\documentclass[3p,twocolumn]{elsarticle}

\usepackage{lineno,hyperref}
\usepackage{amsmath}
\modulolinenumbers[5]

\journal{Astroparticle Physics}

\newcommand{\eh}[1]{\,\mathrm{#1}}

\newcommand{\tin}[1]{_\mathrm{#1}}

\newcommand{\pct}{\eh{\%}}

\newcommand{\che}[1]{}

\renewcommand{\epsilon}{\varepsilon}

\newcommand{\hess}{H.E.S.S.}

\let\seriesfb\bfseries\def\bfseries{\boldmath\seriesfb}
\let\seriesdm\mdseries\def\mdseries{\unboldmath\seriesdm}

\newcommand{\aref}[1]{\ref{#1}}
\newcommand{\tref}[1]{Table~\ref{#1}}
\newcommand{\fref}[1]{Fig.~\ref{#1}}
\newcommand{\eref}[1]{Eq.~\ref{#1}}
\newcommand{\sref}[1]{Sect.~\ref{#1}}

\begin{document}

\begin{frontmatter}

\title{The optimal on-source region size for detections\\with counting-type telescopes}

\author{S.~Klepser\fnref{myfootnote}}
\address{DESY, D-15738 Zeuthen, Germany}
\fntext[myfootnote]{stefan.klepser@desy.de}

%
%

\begin{abstract}
Source detection in counting type experiments such as Cherenkov telescopes
often involves the application of the classical Eq.~17 from the paper
of Li \& Ma (1983) to discrete
on- and off-source regions. The on-source region is typically a circular area with radius $\theta$ in which the signal is expected
to appear with the shape of the instrument
point spread function (PSF). This paper addresses the question of what is
the $\theta$ that maximises the probability
of detection for a given PSF width and background event density.  
In the high count number limit and assuming a Gaussian PSF profile, the
optimum is found to be at $\zeta_\infty^2 \approx 2.51$ times the squared PSF
width $\sigma\tin{PSF39}^2$. While this number is shown to be a good choice in many cases,
a dynamic formula for cases of lower count numbers, which favour larger on-source
regions, is given. The recipe to get to this parametrisation can also be applied
to cases with a non-Gaussian PSF. This result can standardise and simplify
analysis procedures, reduce trials and eliminate the need for experience-based ad hoc cut definitions or expensive
case-by-case Monte Carlo simulations.
\end{abstract}


\end{frontmatter}


\section{Introduction}

Classical on-off detection techniques are still 
widely applied for ground-based gamma-ray observatories like
\hess, MAGIC or VERITAS. In this approach, the event number in a signal
(``on-source'') region is statistically compared to that of an assumedly source-free
background (``off-source'') region. The size of the signal region is defined through
a so-called $\theta^2$-cut, with $\theta$ being the opening angle between
reconstructed gamma-ray direction and source position. Taking into account the
instrument 
point spread function (PSF), the cut
is usually either set to a canonical value of the order of the $68\pct$
containment PSF radius (e.g. in VERITAS \cite{veritas_agn}), or a canonical value of a fixed
efficiency cut ($75\pct$ in MAGIC \cite{magic_upgrade}) or optimised
case-by-case using Monte Carlo (MC) simulations (in H.E.S.S. \cite{hess_crab}). A possible source detection is evaluated using the formula introduced
in Eq.~17 of the famous Li\&Ma paper \cite{lima}.

Although an ad hoc, experience-based choice or a full MC optimisation of the
$\theta^2$ problem can lead to good results, this paper
argues for a simple ma\-thematical solution, which is much easier
and more flexible to apply without computing efforts, and at full
transparency of procedures and trials.

The work in this paper focuses on point-like sources (or sources
with known extension). It neglects the aspects of systematic uncertainties
and Poissonian count numbers, which require other or additional
constraints that can easily be adopted as needed.
 
\section{Nomenclature}

The main parameters that have influence on the number of events in the on-
and off-regions for a given $\theta^2$-cut are the background event density
$n$, the number of photons provided by the source $N\tin{src}$,  and the gamma-ray point
spread function, which in the Gaussian approximation is determined by the
parameter $\sigma\tin{PSF39}$. This Gaussian sigma in two dimensions contains about
$39\pct$ of the signal events\footnote{see \aref{app:formulae} for how to derive it from a $68\pct$
containment radius and more details on the 2D Gaussian calculus used in this
paper}.

In case of a source with known extension, $\sigma\tin{PSF39}$ can simply be replaced by
the source size $\sigma\tin{SRC39}$.
If the PSF (or source extension) is energy dependent, an effective PSF for the considered
energy range has to be computed. For spectral studies, each energy bin might
have its own $\sigma\tin{PSF39}$, in which case the optimal sensitivity
requires one $\theta^2$-cut per energy bin. The following calculations can
thus either be appled to an integral signal or each energy bin of a
spectral study separately.

The calculations are simplified
considerably defining
\begin{equation}\label{eq:defs}\begin{split}
\zeta & = \theta/\sigma\tin{PSF39} \\
\tilde{n}\tin{bkg} & = n\pi\sigma\tin{PSF39}^2
\end{split}\end{equation}
with $\zeta$ being the PSF-scaled $\theta$ and $\tilde{n}\tin{bkg}$ the number of
background events within a circle of radius $\sigma\tin{PSF39}$.
In this case, the 2D Gaussian signal distribution  can be expressed as
\begin{equation}\label{eq:gaussian}
\frac{dN}{d\zeta^2} = \frac{N\tin{src}}{2}\exp(-\zeta^2/2).
\end{equation}
%
%

Locally around the source (within a few $\sigma\tin{PSF39}$), the background of
an instrument with a field of view $\gg\sigma\tin{PSF39}$ is always well-described by an isotropic background
density, which can be extracted a-priori from an off-source $\theta^2$ (or $\zeta^2$)
histogram or skymap. So the expected numbers of excess events and on- and off-events for a given
cut in $\zeta^2$ amount to
\begin{equation}\label{eq:onoff}\begin{split}
N\tin{ex} & = N\tin{src}\,(1-\exp(-\zeta^2/2)) \\
N\tin{off} & = \tilde{n}\tin{bkg}\,\zeta^2\\
N\tin{on} & = N\tin{ex} + N\tin{off}.
\end{split}\end{equation}

\section{Simple case}

An important number can be derived considering the simplified
significance
\begin{equation}\label{eq:simple}
S\tin{simple}(N\tin{ex}, N\tin{off}) = \frac{N\tin{ex}}{\sqrt{2\,
N\tin{off}}},
\end{equation}
which is Eq.~9 from ref.~\cite{lima} assuming $\alpha=1$ and
$N\tin{on} \approx N\tin{off}$ for the denominator. In this case, applying Eqs.~\ref{eq:onoff} leads to 
\begin{equation}\label{eq:simple2}
S\tin{simple}(\zeta^2, N\tin{src},\tilde{n}\tin{bkg}) =
\frac{N\tin{src}}{\sqrt{2\,\tilde{n}\tin{bkg}}}\frac{1-\exp(-\zeta^2/2)}{\sqrt{\zeta^2}}.
\end{equation}
The shape of this function in dependence of the cut value $\zeta^2$ is shown
\fref{fig2} (top left). It has a maximum whose position is invariant against background density
and signal strength, and which can analytically be determined to be
\begin{equation}\label{eq:zetainf} 
\zeta_\infty^2 = -2\mathrm{W}_{-1}\left(\frac{1}{2\sqrt{e}}\right)-1 \approx
2.51 
\end{equation}
where $\mathrm{W}_{-1}(x)$ is the Lambert-W function. A more precise value of
$\zeta_\infty^2$ is shown in \tref{tab1}. 
\section{Li\&Ma case}\label{sec:lima}

\begin{figure*}
\centering
\includegraphics[width=0.49\textwidth]{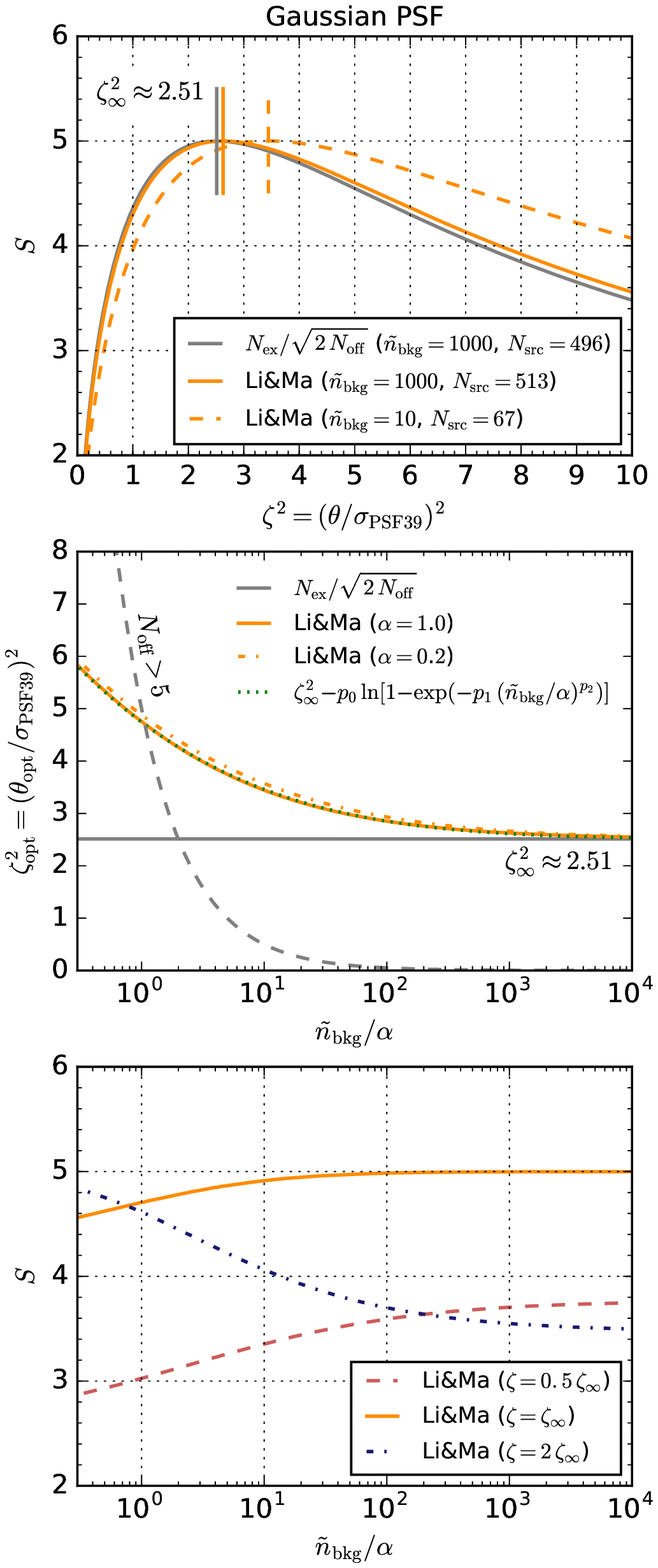}
\includegraphics[width=0.49\textwidth]{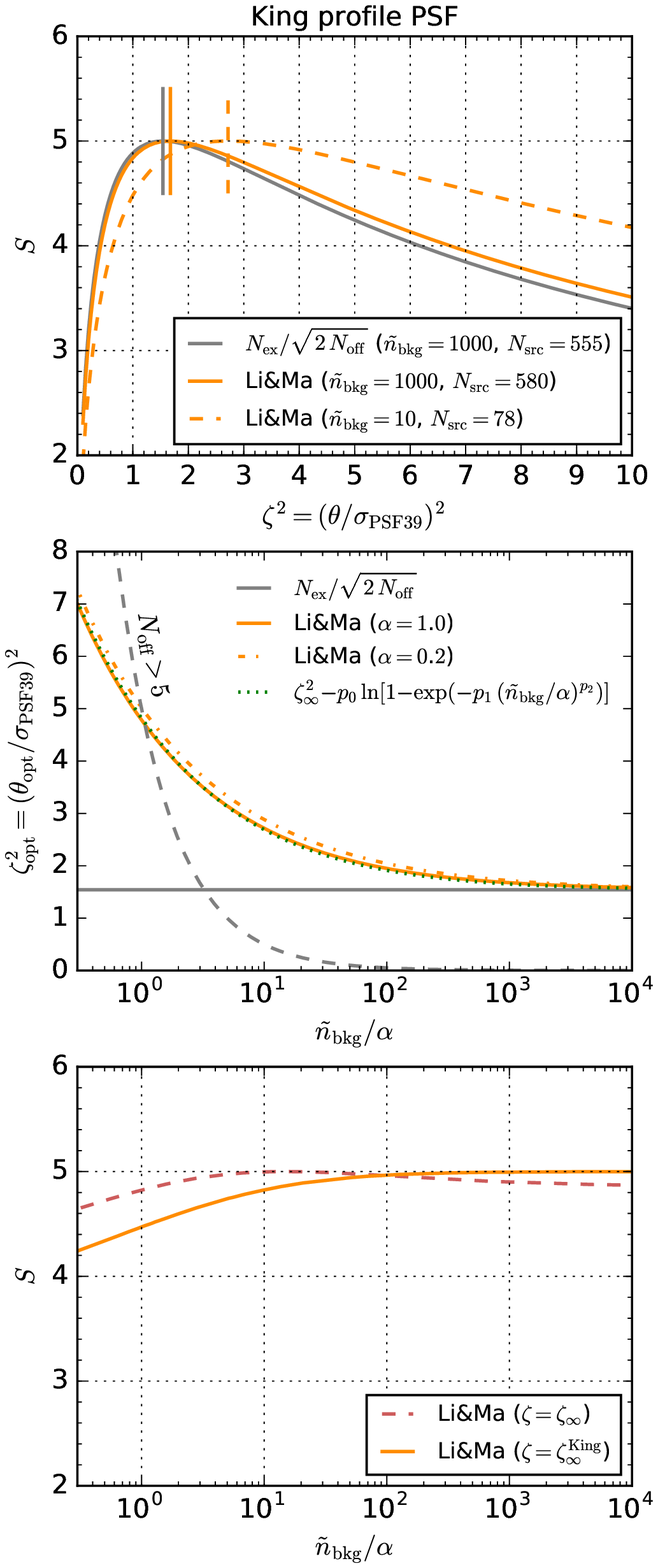}
\caption{
Top left: Examples for the dependency of significance on the on-source region
size. $N\tin{src}$ is
chosen such that it allows for a $5\eh{\sigma}$ detection with the optimal
signal region size cut.
Middle left: Ideal on-source region size as a function of background density in
the Gaussian PSF case.
The line of
$N\tin{off}>5$ gives an indication of where the Li\&Ma formula might lose its
validity in the Poissonian range.
Bottom left: Dependency of significance on background
density for a fixed signal region cut, using $\zeta_\infty$ as the baseline
cut and assuming a too large and too small choice of cut value.
Top right: Same as top left, but using a King profile PSF ($\gamma=2$).
Middle right: Same as middle left, but using a King profile PSF ($\gamma=2$).
Bottom right: 
Same as bottom left, but using a King profile PSF ($\gamma=2$) and either the Gaussian $\zeta_\infty$ as the baseline
cut, or an adjusted fixed value $\zeta_\infty^{\mathrm{King}}$ extracted for
the King profile function (see
straight line in the middle right panel).
\label{fig2}}
\end{figure*}

The Li\&Ma significance depends on the background density and the signal
strength and is therefore slightly more difficult to evaluate. 
As can be seen in \fref{fig2} (top left), it generally needs a slightly higher 
number of source events to get to a given significance value, and favours a
somewhat larger signal region cut.

The complexity can however be reduced if one considers the fact that 
for a source detection, only a signal strength $N\tin{src,5}$ is of interest
that can just actually
lead to a significant detection (typically the canonical $5\eh{\sigma}$). Therefore, the
calculation of this optimum for a given background density can be done in two
dimensions: The maximum of $S\tin{LiMa}(\zeta^2)$ is determined numerically for a given $\tilde{n}\tin{bkg}$ and
$N\tin{src}$, and the latter is increased until $S\tin{LiMa}=5$, resulting
both in $N\tin{src,5}$ and its respective $\zeta\tin{opt}^2$.

Figure~\ref{fig2} (middle left) shows the dependence of the optimal cut on the
background density. Clearly,  in the highly Gaussian regime ($\tilde{n}\tin{bkg}>100$), the optimum cut value
approaches $\zeta_\infty^2$ (therefore the index ``$\infty$''), but cases of low count
numbers favour a somewhat larger cut. This in reverse is equivalent to the
concept outlined in ref.~\cite{hess_crab}, namely that weak sources should be
analysed with tighter cuts (because weak sources require large datasets, i.e.
high background number) and stronger sources with looser cuts. It has to be noted that
the Li\&Ma formula is not valid in the very low-count Poissonian regime,
roughly marked by the line labeled ``$N\tin{off}>5$''.

\begin{figure}
\centering
\includegraphics[width=0.98\columnwidth]{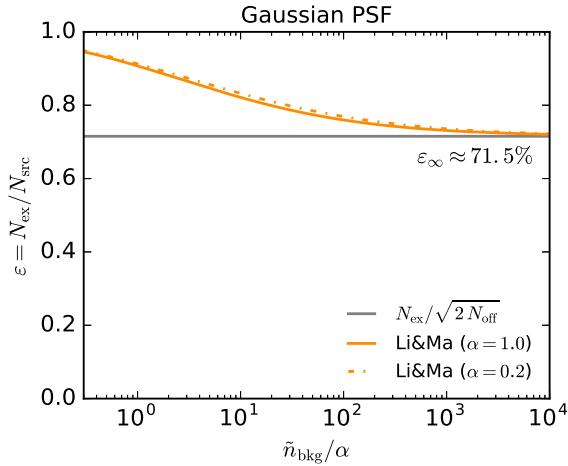}
\caption{
Efficiency of the optimal signal region cut shown in \fref{fig2} (middle left)
for the case of a Gaussian PSF. The efficiency approaches
$\varepsilon_\infty$ at high count numbers, but for Li\&Ma it is favourably higher in the low
statistics regime. \label{eff}}
\end{figure}

The signal event efficiency implied by
the cut is plotted in ~\fref{eff}. In the high
count number limit an efficiency of $71.5\pct$ is approached (proving the
$75\pct$ in MAGIC ref.~\cite{magic_upgrade} likely to be a fair compromise in
many cases).

Figure~\ref{fig2} (bottom left) shows that if the dynamic adjustment of the cut with background density
is replaced by a constant cut, a significance loss of the order of $10\pct$ can be expected in the
low count number case. This is equivalent to a $10\pct$ loss of sensitivity or
a $20\pct$ increase in required observation time.
The dashed and dash-dotted curves furthermore show that if a canonical or weakly motivated
constant cut radius is more than a factor of $2$ away from $\zeta_\infty$,
the sensitivity can even be degraded by a factor of $2$ or more.

In cases where different amounts of on- and  off-exposures are available, the  Li\&Ma
formula offers the application of the parameter $\alpha$, which is the exposure
ratio between the two. Typically, more off- than on-data is available, and
$\alpha$ is smaller than $1$. The dashed-dotted curve in \fref{fig2} (middle 
left) shows that the $\zeta\tin{opt}^2$ curve is almost unaltered if the
background density is scaled to $\tilde{n}\tin{bkg}/\alpha$. 

For simplicity of application, the curve in ~\fref{fig2} (top left) can be parametrised with an
analytical function of the form
\begin{equation}\label{eq:zetapar}
\zeta\tin{opt}^2(\tilde{n}\tin{bkg}, \alpha)=\zeta^2_{\infty} -
p_0\,\ln[1-\exp(-p_1\,(\tilde{n}_{\mathrm{bkg}}/\alpha)^{p_2})]
\end{equation}
The result is shown as a green dotted line in the figure, and the according parameters $p_n$ are
listed in \tref{tab1}.

\begin{table}
\centering
\caption{List of precise numerical constants presented in this paper.}
\label{tab1}
\begin{tabular}{ll}
\hline\hline
Variable & Value \\
\hline
$\zeta_\infty^2$ & 2.51286242 \\
$\zeta_\infty$   & 1.58520106 \\
$\zeta_{\infty,68\pct}$   & 1.04621793 \\
$p_0$ & 160.607603 \\
$p_1$ & 4.28324658 \\
$p_2$ & 0.0789513156 \\
\hline                                   
\end{tabular}
\end{table}

\section{Non-Gaussian point spread functions}\label{sec:non-gaussian}

Although the point spread functions of instruments can usually be approximated
by a Gaussian distribution to some level, the exact distributions are sometimes more complex. Misreconstructed
events can lead to non-Gaussian tails of the PSF. In order to test the
robustness of the above results in these cases, the calculations are repeated
using a so-called King profile
\begin{equation}\label{eq:king}
\frac{dN}{d\zeta'^2} =
\frac{N\tin{src}}{2}(1-1/\gamma)\left(1+\frac{\zeta'^2}{2\,\gamma}\right)^{-\gamma}.
\end{equation}%
This distribution has a tail that is small for large $\gamma$ and gets longer
for $\gamma \rightarrow 1$. $\zeta'$ is defined as 
$\theta/\sigma\tin{King}$, and is related to $\zeta$ like
\begin{equation}\label{eq:kingsigma}
\zeta'^2 = \zeta^2\times2\,\gamma\left[(1-0.39347)^{\frac{1}{1-\gamma}}-1\right].
\end{equation}
The number of excess events in
\eref{eq:onoff} now changes to
\begin{equation}\label{eq:kingexcess}
N\tin{ex}  =
N\tin{src}\left[1-\left(1+\frac{\zeta'^2}{2\,\gamma}\right)^{1-\gamma}\right]
\end{equation}
In this case, the maximum of the corresponding significance function
$S(\zeta)$ depends on $\gamma$ and cannot be derived analytically, even though the
shape is qualitatively similar to the Gaussian case (see \fref{fig2}, top
right, where
a PSF with $\gamma=2$ is taken as an example). The optimum cut in the case of
$S\tin{simple}$, however, is still invariant against $\tilde{n}\tin{bkg}$, and lies somewhat
lower (if calculated w.r.t. $\sigma\tin{PSF39}$, i.e. converting $\zeta'$ to
$\zeta$).

In
the Li\&Ma case, the function of $\zeta\tin{opt}^2$ vs. $\tilde{n}\tin{bkg}$ is
different from the Gaussian PSF case, but can still be
fitted with the parametrisation of \eref{eq:zetapar} (see \fref{fig2},
middle right). Nevertheless, \fref{fig2} (bottom right) shows that,
even if ignoring the fact that a non-Gaussian tail is present, cutting at
$\zeta_\infty$ does not lead to a substantial loss of sensitivity. 
Possibly a too large signal region does lead to a higher background event
number, but in the presence of a tail this disadvantage is partly compensated
by the signal events collected in the tail.

Another consequence of the tail is that the cut efficiency is much
lower in this case (approaching $\sim 50\pct$ for high $\tilde{n}\tin{bkg}$ in the
example of $\gamma=2$). So adjusting the $\theta^2$-cut to a fixed efficiency
$\varepsilon_\infty$ does not universally optimise the significance in cases of a non-Gaussian PSF.



\section{Conclusion}

This paper discusses the question of how large a
signal region of an on-off detection experiment should be in order to optimise
chances of signal detection. It presents an analytical solution for the case
that the point-spread function can to some level be
approximated by a 2D Gaussian profile and that the count numbers are high
enough ($>\mathcal{O}(5)$) such that the Li\&Ma formula can be
applied. It also provides a recipe for non-Gaussian PSF shapes.

The result for the high count number case is $\zeta_\infty^2 \approx 2.51$
(\eref{eq:zetainf}) and answers the question where to cut a $\theta^2$ histogram and also how large a
correlation radius in a skymap kind of analysis should ideally be. It is
equivalent to $\zeta_\infty = 1.585$ times the Gaussian sigma
$\sigma\tin{PSF39}$, or
$1.046$ times the $68\pct$ containment radius\footnote{Which also proves the cut in
the recent VERITAS paper \cite{veritas_agn} to be roughly optimal.}. Precise numbers are given in
\tref{tab1}.

For cases
of lower count numbers ($<100$), slightly larger cuts should be applied using the
parametrisation \eref{eq:zetapar}. This way, a loss of sensitivity
of the order of up to $10\pct$ can be avoided. The formula is only a function of the
background event density divided by the exposure ratio
$\alpha$ and can therefore be determined a-priori or automatically in an
analysis without
a-posteriori adjustments or iterations of cuts.


%
In the case of a non-Gaussian point spread function, \sref{sec:non-gaussian}
shows that the effect of a non-Gaussian tail in the PSF is not
very big, and sticking to the formulae for the Gaussian approximation merely
causes a slight overestimation of the optimal cut, which has a minor impact on the
significance. In extreme cases, though, it is also shown that the formalism
applied in \sref{sec:lima} can easily be adopted for other shapes of the PSF
and can deliver an adjusted fit function \eref{eq:zetapar} if desired.

The currently operating instruments H.E.S.S., MAGIC and VERITAS appear to have
procedures in place that makes them arrive at $\theta^2$-cuts that are relatively close to
optimal. In that sense, this work cannot fundamentally improve their
performance, but rather provides a simple recipe and a reference to avoid an ad hoc
choice of cuts (or cut efficiency), or an overly complicated procedure to define it. 

Although primarily thought to be used in VHE gamma-ray astronomy, the recipe
presented here can also be applied
to other counting-type imagers or in general all problems with the expectation of a signal that can be approximated as
a 2D-Gaussian or 1D-exponential over a flat
background expectation.

In the long run, holistic
likelihood-based analyis frameworks like in refs.~\cite{ctools,
gammapy, generalized_likelihood}, which involve the precise PSF shape in
their fitting, will hopefully outdate the need for the considerations in this
paper and help us to fully exploit the recorded signal events and their
distribution.
 
\section{Acknowledgements}

The author would like to thank S. Ohm for proofreading the manuscript.
This work furthermore made use of
the Python
packages {\tt NumPy}/{\tt SciPy} \citep{numpy, scipy} and {\tt Matplotlib}
\citep{matplotlib}. The mathematical considerations were facilitated using
\url{http://www.wolframalpha.com}. 

\appendix

\section{Calculus around the two-dimensional Gaussian
function}\label{app:formulae}

Since it is hard to find the following summarised in a concise manner
elsewhere, some
basic formulae used in the paper are given here. 

A two-dimensional Gaussian probability density function can be expressed like
\begin{equation}\label{eq:twodgaussian}
\frac{d^2P}{dx\,dy} = \frac{1}{2\pi\sigma^2}\exp\left(-\frac{x^2+y^2}{2\sigma^2}\right)
\end{equation}
In polar coordinates, where
\begin{equation}\label{eq:polar}\begin{split}
x & = \theta\,\cos\phi \\
y & = \theta\,\sin\phi \\
dx\,dy & = \theta\,d\theta\,d\phi,
\end{split}\end{equation}
\eref{eq:twodgaussian} can be expressed as a function of $\theta^2$ only:
\begin{equation}\label{eq:twodgaussianr}
\frac{dP}{d\theta^2} = \frac{1}{2\sigma^2}\exp\left(-\frac{\theta^2}{2\sigma^2}\right)
\end{equation}
The integral of this from $0$ to a radius $\theta\tin{opt}$ is
$P=1-\exp(-\theta\tin{opt}^2/2\sigma^2)$, which makes it very
straight-forward to calculate two-dimensional quantiles or quantile probabilities:
\begin{equation}\label{eq:polar}\begin{split}
P(\theta\leq\sigma)  & = 1-\frac{1}{\sqrt{e}} = 0.3935 \\
P(\theta\leq2\sigma)  & = 1-\frac{1}{e^2} = 0.8647 \\
\sigma\tin{68\%} & = \sigma \sqrt{-2\log(1-0.6827)} = 1.5152\,\sigma \\
\sigma\tin{95\%} & = \sigma \sqrt{-2\log(1-0.9545)} = 2.4860\,\sigma
\end{split}\end{equation}

\section*{References}

\bibliography{mybibfile}

\end{document}